\documentclass{article}
\usepackage{graphicx}

\begin{document}
\rm
\let\sc=\sf
\begin{center}
\LARGE {\bf Dust in the disk winds from young stars as a source of
the circumstellar extinction}
\\
\vspace{10mm} \large {L.V.\,Tambovtseva$^{1}$,
V.P.\,Grinin$^{1,2}$}\\
\vspace{1cm} grinin@gao.spb.ru
\end{center}
\vspace{0.6cm}
\normalsize
1 - Main Astronomical Observatory of RAS Pulkovo, Pulkovskoe shosse 65/1,

196140, St.\,Petersburg, Russia \\
2 - The Sobolev Astronomical Institute of the St.Petersburg University, St.\,Petersburg,
Russia\\

\vspace{10mm} \large {\bf Abstract}. We examine a problem of the
dust grains survival in the disk wind in T Tauri stars (TTSs). For
consideration we choose  the disk wind model described by Garcia
et al. (2001), where a gas component of the wind is heated by an
ambipolar diffusion up to the temperature of the order of 10$^4$
K. It is shown that the dust grains heating due to collisions with
the gas atoms and electrons is inefficient in comparison with
heating by the stellar radiation, and thus, dust survives even in
the hot wind component. Owing to this, the disk wind may be opaque
for the ultraviolet and optical radiation of the star and is
capable to absorb its noticeable fraction. Calculations show that
at the accretion rate $\dot{M_a} = 10^{-8}-10^{-6} M_\odot$ per
year this fraction for TTSs may range from 20\% to 40\% of a total
luminosity of the star correspondingly. This means that the disk
wind in TTSs can play the same role as the puffed inner rim
considered in the modern models of accretion disks. In Herbig Ae
stars (HAEs) inner regions of the disk winds ($r \le 0.5$ AU ) are
free of dust since there dust grains sublimate under the effect of
the radiation of the star. Therefore, in this case a fraction of
the absorbed radiation by the disk wind is significantly less, and
may be compared with the effect of the "puffed-up inner rim" only
at $\dot{M_a} \geq 10^{-6} M_\odot$ yr$^{-1}$. Due to the
structural inhomogeneity of the disk wind its optical depth
towards an observer may be variable resulting in the photometric
activity of the young stars. For the same reason, one can observe
moving shadows from the gas and dust streams with the spiral-like
structure on the highly resolved circumstellar disk images.

\section{Introduction}
The disk wind plays a clue role in the process of removing the
angular momentum excess from accretion disks (Blandford and Payne
1982 (BP82)), and works towards the accretion of the disk matter
towards the star. A physical connection between accretion and
outflow processes is confirmed by a decrease of the outflow
activity with the age of stars, accompanying with a decrease of
accretion rates (Calvet et al. 2000) and a frequency of accretion
disks (see Andr\'{e} et al. (2000); Mundy et al. (2000).)

Most of the modern models (see., e.g., review by Pudritz et al.
(2007)) suggest that the disk wind does not contain the dust.
Nevertheless, already in 1993 Safier (1993a) in his generalized
version of the BP82 self-similar wind model argued that the weakly
ionized wind arising from the accretion disk surface lifts the
dust grains due to their collisions with the neutral atoms, i.e.
the disk wind is dusty. According to Safier, the maximum size of
the grain which are capable to lift with the wind is about of 1
mm.

Apparently, the presence of the dust in the disk wind has to
affect the circumstellar (CS) extinction, the spectral energy
distribution and polarization properties of the young stellar
objects. Based on the results of Safier (1993a), in our previous
papers (Grinin and Tambovtseva 2002; Grinin et al. 2004;
Tambovtseva et al. (2006)) we consider photometric effects
produced by the dusty disk winds in the young binaries. In the
present paper we consider in more details than in the cited papers
an interaction of the dust component of the wind with the
radiation of the star as well with the hot gas for the single TTSs
and HAEs.

\section{Choice of the disk wind model}
\subsection{Observational data}
In more details, observational manifestation of the disk wind was
investigated in TTSs, in whose spectra the wind was responsible
for an origin of some spectral lines, including such forbidden
lines as [O I] $\lambda 6300 \AA,$ [S II] $\lambda 6731 \AA $ and
some others (Solf and B\"{o}hm 1993; Hirth et al. 1994, 1997;
Hartigan et al. 1995). From these line profiles two wind
components have been distinguished: a high velocity component
(HVC) (the jet, 200-400 km/s), which forms in the nearest to the
star regions of the accretion disk, and a low velocity component
(LVC) (5-40 km/s) forming at the periphery regions of the disk
(Kwan and Tademaru 1995). Such separation, however, rather
conventional: studies of the rotational jet velocities revealed an
intermediate wind component provided the poloidal velocities of
the order of 100 km/c (Bacciotti et al. 2000; Lavalley-Fouquet et
al. 2000; Coffey et al. 2004; Woitas et al. 2005).

The presence of the HVC and LVC of the disk wind is especially
spectacular on the images of HH 30. This young star is surrounded
by the CS disk seen nearly edge-on; as a result, in the visible
part of the spectrum a direct radiation of the star is completely
absorbed by the disk. Visual images of HH 30 obtained with the
Hubble Space Telescope distinctly demonstrate highly collimated
jet propagating in the direction perpendicular to the disk plane
with the mean velocity of about 300 km/s. Observations in the CO
molecule lines showed (Pety et al. 2006) that in the same
direction a slow biconical outflow was observed with the typical
velocity of about 12 km/s and the open angle of about 30 degrees.
The mass loss rate in the biconical outflow estimated by these
authors was $\sim 6.3 \cdot 10^{-8} M_\odot$ per year, while that
in the jet was less by about two orders of magnitude ($\sim
10^{-9} M_\odot$ per year). Hence it follows, that a bulk of the
kinetic energy of the disk wind ($L_w \approx 6\cdot 10^{31}$
erg/s) comes to acceleration of the matter in the narrow
collimated jet, while the main mass loss falls to the low velocity
wind component. Such a conclusion agrees well with results of the
numerical MHD simulations (Goodson et al. 1999).

\subsection{Theoretical models}
According to Blandford and Payne (1982), processes of acceleration
and tap of the matter in the disk wind are governed by the
magnetic field of the accretion disk. If magnetic fields lines
threading a thin rotating disk make an angle $\vartheta$ with the
symmetry axis of the disk $ \ge \vartheta_0$ ($ \vartheta_0 =
30^\circ$), then the disk matter will be accelerated under the
effect of the Lorentz force and the magneto-centrifugally driven
wind can be launched. If the disk is threaded by the open field
lines from some internal radius $r_i$ to some external radius
$r_e$, then one has a typical scheme of the extended disk wind for
which different self-similar solutions have been obtained with the
help of magnetohydrodynamics (K\"{o}nigl 1989; Wardle and
K\"{o}nigl 1993; Ferreira and  Pelletier 1995; Ferreira 1997;
Casse and Ferreira 2000; Ferreira and Casse 2004)\footnote{There
is so-called X-wind model where the wind is launched from one
annulus located in the inner disk (Shu et al. 1994; Shang et al.
2002); in this model the disk does not possess the large scale
magnetic field, and the main role is assigned to the magnetosphere
of the star. In the framework of this model some modern
observational facts are not reproduced (see., e.g. Ferreira et al.
2006, Coffey et al. 2004).}. There is the parameter $\xi = d \log
\dot M_{a}(\varpi)/d \log\varpi$ which is a measure of the disc
ejection efficiency and regulates relation between low-velocity
and high velocity wind components, (here $\dot M_{a}$ is the
accretion rate, $\varpi$ the distance from the disk symmetry
axis). Calculations show that the best agreement with the observed
parameters of the forbidden lines in the spectra of TTSs occurs at
$\xi \approx$ 0.007-0.01 (Cabrit et al. (1999), Garcia et al.
2001a). Below in calculations we use the model "A" from the paper
by Garcia et al. (2001a) where the parameter $\xi$ = 0.01.

\subsection{The model of the dust mixture}
During the evolution, the dust component of the protoplanetary
disk undergoes essential changes: the dust grains grow and
gradually settle towards the disk midplane (Safronov 1972;
Weidenschilling 2000). Further, they form solids and
planetezimals. However, in the surface layers of the disk small
grains of an approximately original (i.e. interstellar) chemical
composition persist during a long time. Results of photometric
observations of UX Ori type stars testify this. A violent
photometric activity of these stars is caused by the changes in
the CS extinction due to the small inclination of their CS disks
relatively to the line of sight (see review of Grinin (2000) and
papers cited there). The data on the selective CS absorbtion,
which is observed in these stars during their fading show that the
reddening law is close to the interstellar one (see, e.g., Pugach
2004). Since the disk wind starts from the surface of the CS disk
we operate below with the MRN mixture (Mathis et al. 1977). Along
with this, we also consider the dust grains with the radius equal
to $a$ = 0.1 $\mu$m which provide the reddening law close to that
given by the MRN mixture.

\section{The dust survival in the gas component of the wind}
As it was shown by Safier (1993a,b), an ambipolar diffusion (the
ion-neutral drift) is an important source of the gas heating in
the disk wind of TTSs. Under the effect of this mechanism the
accelerated gas is heated up to the temperature of about 10$^4$ K.
In the wind regions nearest to the star one can also expect an
essential contribution from the X-ray radiation to the gas heating
(see., e.g. Glassgold et al. 2000), whose major part originates in
the shocks during infall of the accretion gas onto the star. The
question arises: can the dust grains survive contacting with the
heated gas?

\subsection{Collisions with the gas atoms. Thermal effect}
When the gas particles collide with the dust grains, a part of
their kinetic energy converts into the heat resulting in the dust
heating. An efficiency of such a process depends on the sort of
particles (atoms, ions, electrons) and is determined by the
relation (see, e.g. Draine 1981):

\begin{equation}
Q_{coll} = \pi a^2 \sum_i n_i(\frac{8kT_i}{\pi m_i})^{1/2}\,2kT_i
<\alpha_i>
\end{equation}

Here $n_i, m_i$ and $T_i$ are number densities, masses and kinetic
temperatures of gas species $i$, $a$ is the radius of the dust
grain, $<\alpha_i>$ the mean fraction of the kinetic energy which
is converted to heat when a particle $i$ impacts the grain, $k$
the Boltzmann's constant.

Let us consider an efficiency of this mechanism in comparison with
heating of the dust grains by the stellar radiation. An energy
absorbed by the dust grain can be written as
\begin{equation} Q_{*}
= \frac{\pi a^2}{4\pi r^2} \int_0^\infty\,L_*(\lambda)\,
Q_{abs}(\lambda)\,d\lambda,
\end{equation}
where $r$ is the spherical radius, $L_*(\lambda)$ the luminosity
of the star at the given wavelength $\lambda$, $Q_{abs}$ the
absorption efficiency factor for the grains of the given radius
and chemical composition.

The ratios of $Q_*/Q_{coll}$ for two sorts of the dust grains
(graphite and astronomical silicate) are shown in Figs. 1 and 2.
The grain radius is equal to 0.1 $\mu$m. The effective temperature
$T_{eff}$ and the radius $R_*$ of the star are equal to 4000 K and
$2.5 R_\odot$ respectively. The spectral energy distribution of
the star is described by the Planck function. Calculations are
made for two streamlines: the innermost one with the start
coordinate in the disk plane $\varpi_0$ = 0.1 AU and an outer
streamline with $\varpi_0$ = 1 AU. The medium was assumed to be
optically thin for the stellar radiation. Optical characteristics
of the dust were calculated with the Mie theory. The optic
constants were taken from the paper by Draine (1985).

One can see that in the both cases the dust heating due to
collisions with atoms and free electrons in the disk wind is
negligible in comparison with that by the radiation of the star.
Only in those wind regions where the radiation of the star is
strongly diluted due to the absorbtion by the dust component of
the wind, heating due to collisions may be dominant. But even
there cooling of the dust due to the radiation is an efficient
process and the grain temperature is far from the sublimation one.
As shown by Safier (1993a), the opposite process (the gas cooling
by the dust) plays an important role in the base of the wind but
in the higher wind layers is less effective than the adiabatic
cooling.
\begin{figure}
\centering
\includegraphics[angle = -90, width=12cm]{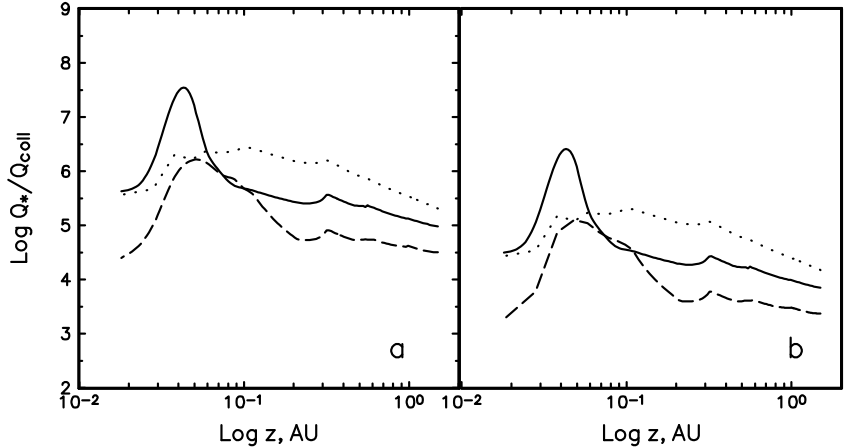}
\caption{A ratio of the heat power by the radiation of the star to
that by dust-gas collisions along the streamline with the anchor
at $\varpi_0$ = 0.1 AU, the grain radius is 0.1 $\mu$m, (a)
graphite, (b) astrosilicate. The accretion rate is equal to
$10^{-6} M_\odot$ per yr. (dashed line), $10^{-7} M_\odot$ per yr.
(solid line) and $10^{-8} M_\odot$ per yr. (dots).}
\end{figure}
\begin{figure}
\centering
\includegraphics[angle = -90, width=12cm]{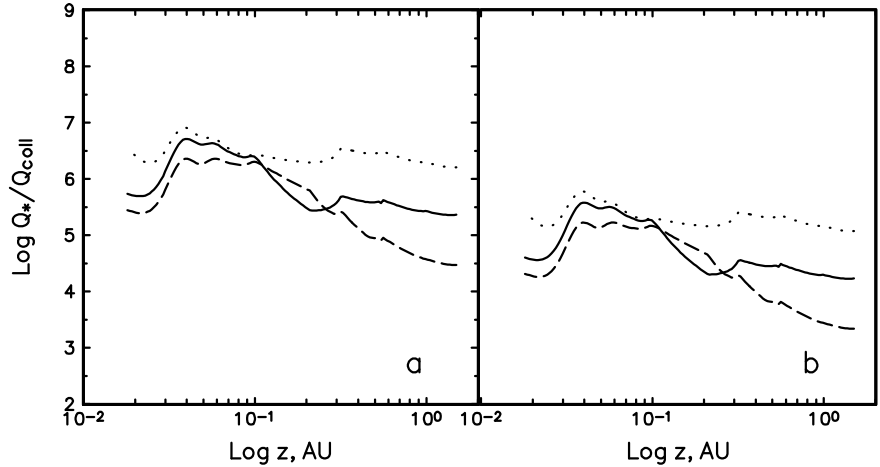}
\caption{The same as in Fig. 1 but for the streamline with
$\varpi_0$ = 1 AU}
\end{figure}

Thus, the hot gas in the disk wind and the cold dust can exist in
the same regions of the medium, and there is no any paradox or
contravention of the thermal dynamics laws. The reason is that the
disk wind is transparent or semi-transparent for the thermal
radiation of the dust grains and, therefore is not a close system
in the thermodynamical sense.

\subsection{Dust sputtering and sublimation}
A dust grain can dissipate in the \textit{sputtering} process,
when molecules are ejected from the grain after the latter
collides with gas particles; this leads to the destruction of the
grain. In this case the mass loss by the dust particle is

\begin{equation}
    \frac{dm}{dt}=-m_s\sum_iN_iY_i,
\end{equation}
where $N_i$ is the number of particles of species $i$ impacting
the dust grain in the unit of time, $m_s=\mu_sm_H$ the mass of the
molecule leaving the grain, $\mu_s$ its molecular weight, $Y_i$
sputtering yield, i.e. the number of molecules released after
impact by the particles of the given species. The value of $Y_i$
strongly depends on the energy of colliding particles (Draine and
Salpeter 1979). According to these authors, at the energy of the
incident atoms of about 1 ev , the sputtering yield is equal to
zero both for silicate and graphite. The same is valid for the
"chemical" sputtering (Draine 1979).

Thus, the main process affected the dust survival in the disk wind
is the sublimation of the dust in the radiation field of the star.
As mentioned above, Safier (1993a) and Gracia et al. (2001)
determined the dust sublimation zone in the disk winds of TTSs:
this is a region extended approximately to 0.1 AU from the central
source. In the case of HAEs the sublimation zone is greater. From
our calculations it is about 0.5 AU for the model adopted below.
Therefore, the inner regions of the disk wind in HAEs are free of
dust.

\section{Disk wind and circumstellar extinction}
Assuming that the disk wind has an axial symmetry we estimated the
portion of the total luminosity of the star which can be absorbed
and scattered by the dust component of the wind, and the disk
inclination angles under which the wind becomes transparent for
the radiation of the star. The first of these parameters (we call
it as screening coefficient) is determined as follows:
\begin{equation}
\delta = \frac{1}{L_*}\,\int_0^\infty
L_*(\lambda)\,d\lambda\int_0^{\pi/2} (1 -
e^{-\tau_0(\lambda,\theta)})\sin{\theta}\,d\theta
\end{equation}
Here $L_*$ is a bolometric luminosity of the star (as above we
assume the Planck spectrum), $\theta$ an angle between an
arbitrary radius-vector $\vec r$ and the symmetry axis of the
disk; $\tau_0(\lambda,\theta)$ the optical depth of the disk wind
at the wavelength $\lambda$ in the $\vec r$ direction.

Calculations of $\tau_0$ have been made for the MRN mixture at
dust to gas ratio 1:100 that is typical for the interstellar
medium. The gas density distribution in the TTS's wind was taken
as mentioned above. The same model was used for the disk wind in
HAEs. For this purpose we used scaling given in the Garcia et al
(2001a) (relations (9)) connecting the disk wind parameters with
the mass of the star and the accretion rate. For HAEs we adopt
$M_* = 2.5 M_\odot$. Two other parameters needed for calculation
of the sublimation radius are the stellar luminosity ($L_* = 50
L_\odot$) and the effective temperature ($T_{ef} = 9000$ K).

Results of the $\delta$ calculations for different values of the
accretion rate in the range of $\dot M_a$ = $10^{-9}$ - $10^{-6}
M_\odot$ per year are shown in Fig. 3. It is seen that for TTSs
the value of $\delta$ changes in the range from about 0.1 at $\dot
M_a$ = $10^{-9} M_\odot$ yr$^{-1}$ to $\approx$ 0.4 at $M_a =
10^{-6} M_\odot$ yr$^{-1}$. This means that at $M_a \geq 10^{-8}
M_\odot$ yr$^{-1}$ \emph{the dust component of the disk wind can
absorb and scatter a noticeable fraction of the stellar radiation
producing thus an expanding shadow zone in the adjacent regions of
the CS disk}.

Figure 3 presents also results of the analogues calculations for
graphite - silicate mixture with proportions as in Draine and Lee
(1984) and with the fixed radius ($a$ = 0.1 $\mu$m). Such a
mono-dispersed mixture will be used in the further paper in
simulations of the infrared radiation of the disk wind, since in
the visual and near infrared regions of the spectrum it has
optical characteristics very close to that of the MRN mixture. It
is indirectly confirmed by Fig. 3: it is seen that this mixture
provides almost the same screening effect by the disk wind as the
MRN mixture.

Calculations of the thermal balance of the graphite and silicate
grains with the radius $a$ = 0.1 $\mu$m showed that for Herbig Ae
stars with adopted $L_*$ and $T_{ef}$, the sublimation radius is
equal to 0.35 AU for graphite particles and 0.75 AU for silicate
ones. Taking this into account we calculated optical depths in the
disk wind $\tau_0$ and coefficient of screening $\delta$ for HAEs.
It is seen from Fig. 3 that an absence of dust in the inner part
of the disk wind in HAEs notably decreases a solid angle within
which the radiation of the star can be absorbed and scattered by
the dust component of the wind: a maximum value of $\delta$ (at
$\dot M_a$ = 10$^{-6} M_\odot$ yr$^{-1}$) is about 0.15; this is
less by $\sim$ factor of 3 than that for TTSs at the same $\dot
M_a$ but comparable with the effect produced by the puffed-up
inner rim in the dust sublimation zone (Natta et al. 2001).

\begin{figure*}
\centering
\includegraphics[angle = -90, width=9cm]{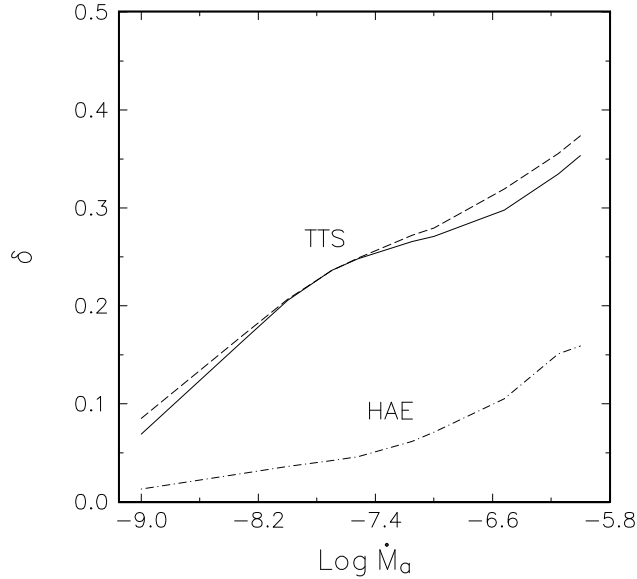}
\caption[]{The coefficient of screening of the stellar radiation
by the dusty disk wind vs. $\dot M_a$. Solid line: the MRN
mixture, dashed line: the mono-dispersed mixture with radius 0.1
$\mu$m (both curves relate to the disk wind in TTSs),
dashed-dotted line: the mono-dispersed mixture with the radius 0.1
$\mu$m for the disk wind in HAEs.}
\end{figure*}
Figure 4 shows angles $\theta_1$ between the disk plane and the
line of sight at which the optical depth of the disk wind $\tau$
is equal to unity at wavelengths $\lambda$ = 0.5 and 0.1 $\mu$m.
The former is close to the maximum of the V - band path, the
latter is close to the wavelength of the $L_\alpha$ - line which
plays an important role in the energetics of the ultraviolet
spectra of the young stars. Calculations are fulfilled for MRN
mixture. It is seen that in TTSs the angle $\theta_1$ at which
$\tau_{\lambda_{0.5}}$ = 1 ranges from 8 to 37 degrees depending
on the accretion rate. In HAEs the corresponding values of
$\theta_1$ are notably less because of the existence of the inner
region in the disk wind free of dust.

At $\lambda = 0.1 \mu$m the extinction coefficient of the MRN
mixture is three times greater than that at $\lambda = 0.5 \mu$m.
As a result, the angle $\theta_1$ corresponded to
$\tau_{\lambda_{0.1}}=1$ increases. For TTSs it reaches 12 - 45
degrees at accretion rates $10^{-9} - 10^{-6}M_\odot$ per year
respectively. Calculated utmost angles show under which
inclinations of the CS disks to the line-of-sight one can see
ultraviolet and optical spectra of the young stars undisturbed by
the absorption in the disk wind.

\begin{figure*}
\centering
\includegraphics[angle = -90, width=15cm]{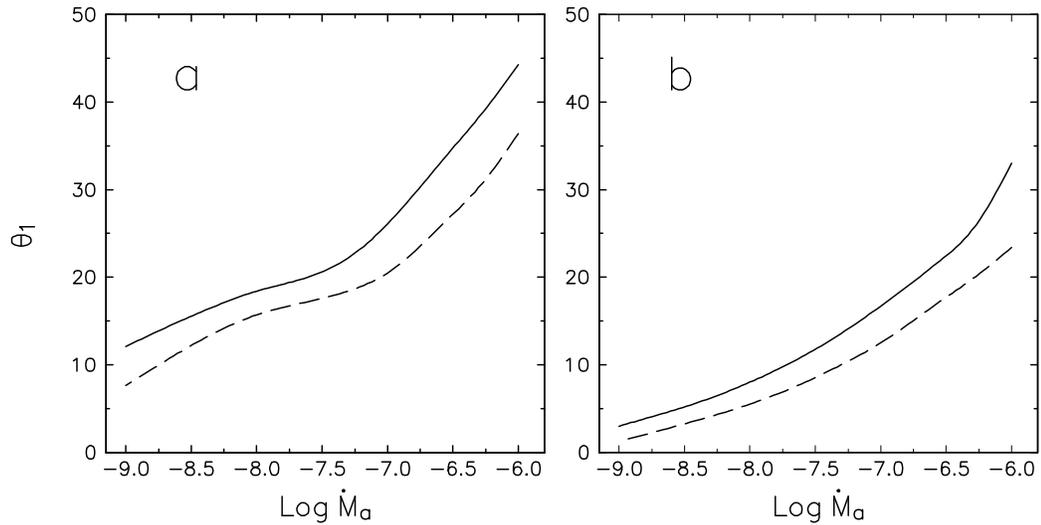}
\caption[]{An angle between the disk plane and the line of sight
at which the optical depth of the disk wind is equal to unity at
the wavelength $\lambda = 0.5\mu$m (solid) and 0.1$\mu$m (dashed)
\textbf{a)} for TTSs, \textbf{b)} - for HAEs. See the text for
details.}
\end{figure*}

\begin{figure*}
\centering
\includegraphics[width=12cm]{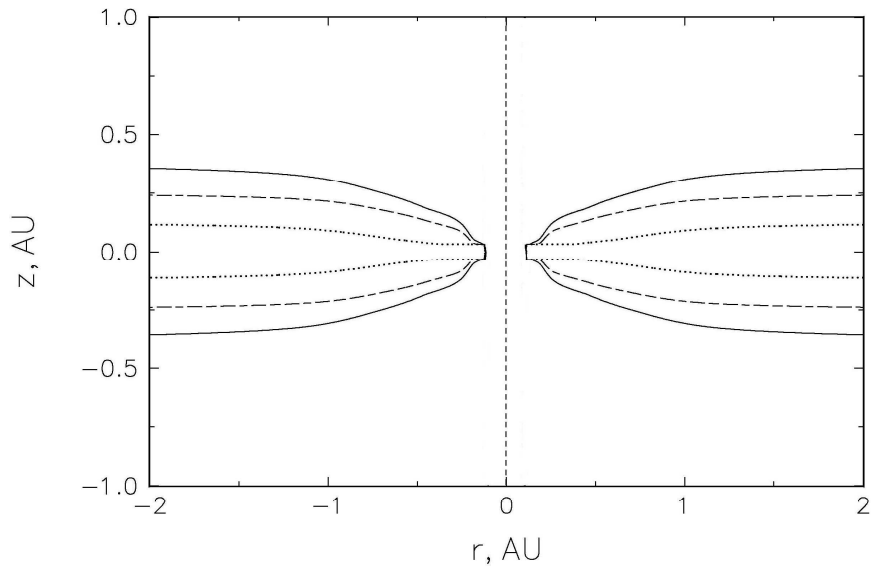}
\caption[]{Contours of the TTSs' disk wind at $\tau_\bot$ = 1 at
the wavelengths $0.1\mu$m (solid line) and 0.5$mu$m (dashed line).
The accretion rate $\dot{M_a}=10^{-6}M_\odot$ per year.}
\end{figure*}

Figure 5 shows contours of the disk wind in TTSs calculated under
the condition that the vertical optical depth $\tau_\bot$ measured
from the disk surface inwards is equal to unity. Calculations have
been made for three wavelengths: 0.1, 0.5 and 3 $\mu$m for the
wind model with $\dot{M_a}=10^{-6}M_\odot$ per year and MRN dust.
The wavelength 3$\mu$m corresponds to the effective wavelength of
the infrared radiation originated in the regions of the CS disk
nearest to the star. At this $\lambda$ a mayor part of the disk
wind is transparent for the radiation. Therefore, its boundary at
3 $\mu$m is close to the surface of the standard flared disk
(e.g., Kenyon and Hartmann 1987; Dullemond and Natta 2003)
determined as $H/r \approx 0.1,$ where $H$ is the scaling height
at the distance $r$. In the visual and especially in the
ultraviolet spectrum regions the effective optical depth of the
disk wind increases, and one has to take into account this
circumstance when modelling the spectral lines arising in the
dense wind layers (such as, for example, ultraviolet molecules
H$_2$).

\section{Discussion}
Thus, we showed that in the case of T Tauri stars the disk wind
can absorb and scatter the radiation of the star within a rather
large segment of the solid angle 4$\pi$ in the wide range of the
accretion rates $\dot{M_a}= 10^{-8} - 10^{-6}M_\odot$ yr$^{-1}$.
This means that the disk wind, in fact, is capable to play the
same role as the puffed-up inner rim in the dust sublimation zone
of the accretion disk (Natta et al. 2001). This inner rim screens
the adjoining regions of the accretion disk from the direct
stellar radiation (Dullemond et al. 2001), and under the certain
inclination angles of the disk to the line of sight may be a
source of the variable CS extinction (Dullemond et al. 2003). The
dust component of the disk wind is able to produce the same
effect.

In the case of Herbig Ae stars a screening effect produced by the
dust component of the disk wind at the same range of accretion
rates is significantly less than that for TTSs. Therefore, in HAEs
the contribution of the disk wind into the thermal radiation of
the CS dust can be comparable with that from the inner rim only at
the high values of the accretion rate $\geq 10^{-6}M_\odot$
yr$^{-1}$.

It should be noted that we considered optical properties of the
dust component of the wind using the model of Garcia et al (2001)
with $\xi = 0.01$. In the disk wind theory this important
parameter governs an efficiency of the magneto centrifugal
mechanism of the gas acceleration. In particular, a growth in
$\xi$ leads to an increase of the mass loss rate in the disk wind
as well to a decrease of the terminal velocity of the wind (Garcia
et al. 2001). Both effects works in the same direction: they
increase the density of the matter in the wind. Therefore in the
models with the large $\xi$ the disk wind has to be more opaque on
dust in comparison with the model considered above.

\subsection{Structural disk wind and variable circumstellar
extinction} Basing on the existing disk wind models we suggested
that the wind possesses an axial symmetry and azimuthal
homogeneity. In fact, this is a model simplification, and in
reality, it seems hardly feasible. In conditions of supersonic
turbulent motions the disk wind cannot be a continuous outflow in
the hydrodynamical sense. It has to consist of an aggregate of gas
and dust streams of the different power arising from the disk
surface. In such a case the filling factor $q$ would be one of the
wind parameters. It implies a fraction of the wind volume filled
in with streams of the matter. Now we can only note that $q$ has
to be less than unity.

Thus, under the real conditions, the column density of the dust
along the line of sight passed through the disk wind, may be a
complex function of time. It may fluctuate due to the motion of
the gas and dust streams. Besides its changes can reveal
quasi-periods caused by the repeated intersection of the line of
sight by the same dominant wind stream. Note, that the
quasi-periods in the brightness changes have been really observed
in some UX Ori type stars (see e.g., Shevchenko et al. 1993). The
rotation of the nonhomogeneous disk wind could be the reason of
the spectral variability of some young stars (e.g. Kozlova et al.
2003).

Changes in the CS extinction may vary not only the radiation flux
coming to the observer directly from the star, but the radiation
flux from that region of the disk which is illuminated by the star
through the disk wind. Shadows from the wind in this part of the
CS disk have to move along the disk following the motion of the
gas and dust streams. Since these streams looks like spinning-up
spirals, their shadows projected on the disk have to be also
spiral-like. Detection and investigation of such moving shadows on
the images of the CS disks would be important for the theory of
the disk winds.

{\bf HH 30}. It is likely that namely such a mechanism of the
variability is realized in the case of HH 30. Comparison of the
images of this object obtained in the different time with the
Hubble Space Telescope showed that they are variable. Both the
type of the object's asymmetry  and the integral flux from it were
variable (Burrows et al. 1996; Stapelfeldt et al. 1999, Wood et
al. 2000). Wood and Whitney (1998) supposed that changes in
conditions of illumination of the CS disk by the spotted rotating
star could be the reason of the HH 30 variability. However, new
data on the variability of the object (Watson and Stapelfeldt
2004, 2007) did not confirm the presence of the period connected
with the rotation of the spotted star. According to these authors,
variability of HH 30 has a more complex character and caused by
changes in the CS extinction in the inner regions of the disk. A
structural disk wind consisted of the separate gas and dust
streams starting from the surface of the CS disk corresponds well
to this role.

{\bf RW Aur}. Another example of the young star whose variability
is difficult to explain without appealing to a hypothesis about a
dusty disk wind is the classical TTS RW Aur. This star relates to
the most studied young stars. It is characterized with a large
amplitude photometric activity (Herbst et al. 1994) and a complex
type of variability of the emission line profiles and intensities
(Petrov et al. 2001; Alencar et al. 2005). Recently Petrov and
Kozak (2007) analyzed in detail a long-term series of the spectral
and photometric observations of RW Aur and showed that there is a
correlation in variations in the emission lines with the different
excitation potential, which can be explained only if to assume
that spectral variability is due to screening the emission region
by the CS dust clouds. It is known from observations that the
symmetry axis of the RW Aur's CS disk is inclined to the line of
sight under 46 $\pm 3^\circ$ (this angle was derived very
accurately with the help of the radial and space (a projection on
the sky plane) velocities of the moving details in the optical jet
(Lopez-Martin et al. 2003)). Under such an inclination disk cannot
screen the star even if to take into account the rim in the
sublimation zone. Therefore, an appearance of the dust on the line
of sight (and hence, on the high latitudes in the star's
coordinate system) Petrov and Kozak connected with the dust
fragments of the disk wind.

An applicability of the theory of the dusty disk winds is not
limited by examples given above. The calculations show (Grinin and
Tambovtseva 2002; Tambovtseva et al. 2006) that the photometric
effects caused by the dust component of the disk wind can be
observed in the young binaries. In particular, obscuration by the
extended disk win could cause abnormally long lasting eclipses
observed in some binaries.

\section{Conclusion}
Let us briefly summarize results of the analysis given above.

1. Basing on the disk wind model described by Garcia et al. (2001)
we showed that the dust grains carrying away by the gas component
of the wind survive being in the contact with the hot ($10^4$ K)
gas.

2. The range of the solid angles which is covered by the part of
the wind opaque by dust depends on the accretion rate and the
luminosity of the star, and for TTSs may amount a noticeable
fraction of the full solid angle 4$\pi$. This means that the disk
wind can notably contribute both to the scattered radiation at the
optical and ultraviolet wavelengths, and to the infrared excesses
of the radiation of T Tauri stars.

3. Conditions of the disk wind formation are such that it cannot
be a continuous axially-symmetric outflow; it is rather an
agglomerate of the gas and dust streams started from those points
of the circumstellar disk where the conditions for the matter
acceleration by the magnetic field are most favorable. A motion of
the matter in the disk wind results in the variations of the dust
column density on the line of sight. Therefore, under certain
inclination of the disk to the line of sight the gas and dust
streams of the disk wind can cause the variable CS extinction
resulting in the photometric activity of the young stars. For the
same reason one can see moving shadows on the CS disks images
caused by gas and dust streams arising from the disk surface.

4. Herbig Ae stars have the sublimation radius at about 0.5 AU
from the central source. As a result, the inner densest part of
the disk wind is free of dust, and the effective solid angle
within which the dust wind component can interact with the
radiation of the star is small compared to 4$\pi$. Nevertheless,
even in such a case a periphery region of the wind may be a source
of the variable CS extinction, responsible for the photometric
activity of UX Ori type stars. Therefore, dense in time
photometrical monitoring of these stars may give a valuable
information on the disk wind structure in the acceleration zone in
the close vicinity to the surface of the accretion disk.

We are grateful to A. K\"{o}nigl for useful discussion of the
results of this work and valuable comments. The work is supported
by the Program of the Presidium of RAS "Origin and evolution of
stars and Galaxies", INTAS grant N 03-51-6311 and grant
NS-8542.2006.2.

\begin{center}
\LARGE {\bf References}
\end{center}
S.H.P. Alencar, G. Basri, L. Hartmann, N. Calvet, 2005, Astron. Astrophys.
\textbf{440}, 595\\
P. Andr\'{e}, D. Ward-Thompson, M. Barsony, 2000,
\textit{Protostars and Planets IV}, (eds. V. Mannings, A.P. Boss,
S.S. Russel, The University of Arizona Press, Tucson), p. 59.\\
F. Bacciotti, R. Mundt, T. P. Ray et al. 2000, Astrophys. J. \textbf{537}, L49\\
R. D. Blandford and D.G. Payne, 1982, MNRAS, \textbf{199}, 883\\
C. J. Burrows, K. R. Stapelfeldt, A.M. Watson et al. 1996,
Astrophys.J., \textbf{473}, 437\\
S. Cabrit, J. Ferreira, A.C. Raga, 1999, Astron. Astrophys., \textbf{343}, L61 \\
N. Calvet, L. Hartmann, S.E. Storm, 2000, \textit{Protostars and
Planets IV},(Eds. V. Mannings, A.P. Boss, S.S. Russel,
The University of Arizona Press, Tucson), p. 377\\
F. Casse and J. Ferreira, 2000, Astron. Astrophys. \textbf{353}, 1115\\
D. Coffey, F. Bacciotti, J. Woitas, T.P. Ray, and J.
Eisl\"{o}ffel, 2004, Astrophys. J. \textbf{604}, 758 \\
B.T. Draine, 1979, Astrophys. J., \textbf{230}, 106\\
B.T. Draine, 1981, Astrophys. J., \textbf{245}, 880\\
B.T. Draine, 1985, Astrophys. J. Suppl. Ser., \textbf{57}, 587\\
B.T. Draine and E.E. Salpeter, 1979, Astrophys. J, \textbf{231}, 77\\
B.T. Draine and E.E. Salpeter, 1979, Astrophys. J. \textbf{231}, 77\\
B.T. Draine, H.M. Lee, 1984, Astrophys. J. \textbf{285}, 89,\\
C.P. Dullemond and A. Natta, 2003, Astron. Astrophys.
\textbf{408}, 161\\
C.P. Dullemond, C. Dominik, and A. Natta, 2001,
Astrophys. J., \textbf{560}, 957\\
C.P. Dullemond, M.E. van den Ancker, B. Acke, and R. van Boekel
2003, Astrophys. J., \textbf{594}, L47\\
J. Ferreira, 1997, Astron. Astrophys., \textbf{319}, 340\\
J. Ferreira and G. Pelletier, 1995, Astron. Astrophys. \textbf{295}, 807\\
J. Ferreira and F. Casse, 2004, Astrophys. J., \textbf{601}, L139\\
J. Ferreira, C. Dougados, S. Cabrit, 2006, Astron. Astrophys. \textbf{453}, 785\\
P.J.V. Garcia, J. Ferreira, S. Cabrit, and L. Binette, 2001a,
Astron. Astrophys. \textbf{377}, 589\\
P.J.V. Garcia,  S. Cabrit, J. Ferreira,, and L. Binette, 2001b,
Astron. Astrophys. \textbf{377}, 609\\
A. E. Glassgold, E.D. Feigelson, T. Montmerle, 2000,
\textit{Protostars and Planets IV}, (eds. V. Mannings, A.P. Boss,
S.S. Russel, The University of Arizona Press, Tucson, p. 457.\\
V.P. Grinin, 2000, \textit{Disks, Planetesimals, and Planets},
(Eds. F. Garzon, C. Eiroa, D. de Winter, and T. J. Mahoney, ASP
Conference Proceedings, Vol. 219, Astronomical Society of the Pacific), p.216\\
V. P. Grinin and L..V Tambovtseva, 2002, Astron. Letters \textbf{28}, 601\\
V.P.Grinin, L.V. Tambovtseva, N.Ya. Sotnikova, 2004, Astron. Lett. \textbf{30}, 694\\
A.P. Goodson, K.-H. B\"{o}hm, and R.M. Winglee, 1999, Astrophys. J.  \textbf{524}, 142\\
P. Hartigan, S. Edwards, L. Ghandour, 1995, Astrophys. J. \textbf{452}, 736\\
W. Herbst, D.K. Herbst, and E.J. Grossman, 1994, Astron. J. \textbf{108}, 1906\\
G. A. Hirth, R. Mundt, J. Solf, and T.P. Ray, 1994, Astrophys. J., \textbf{427}, L99\\
G. A. Hirth, R. Mundt, J. Solf, 1997, Astron . Astrophys. Suppl. Ser., \textbf{126}, 437\\
S.J. Kenyon and L. Hartmann, 1987, ApJ \textbf{323}, 714\\
A. K\"{o}nigl, 1989, Astrophys. J. \textbf{342}, 208\\
O.V. Kozlova, V.P.Grinin, G.A. Chuntonov, 2003, Astrophysics,
\textbf{46}, 265\\
J. Kwan and E. Tademaru, 1995, Astrophys. J. \textbf{454}, 382\\
C. Lavalley-Fouquet, S. Cabrit, and C. Dougados, 2000, Astron.
Astrophys. \textbf{356}, L41\\
L. L\'{o}pez-Martin, S. Cabrit and C. Dougados, 2003, A\&A, 405, L1\\
J.M. Mathis, W. Rumpl, and K. H. Nordsieckm 1997, Astrophys. J.
\textbf{217}, 425\\
L.G. Mundy, L.W. Looney, W.J. Welch, 2000, \textit{Protostars and
Planets IV}, (eds. V. Mannings, A.P. Boss, S.S. Russel,
The University of Arizona Press, Tucson), p. 355\\
A. Natta, T. Prusti, R. Neri et al. 2001, Astron. Astrophys., {\bf
371}, 186 \\
P.P. Petrov, G.F. Gahm, J.F. Gameiro, et al. 2001, Astron.
Astrophys. \textbf{369}, 993\\
P.P. Petrov P.P and B.S. Kozak, 2007, Astron. Rep. \textbf{51}, 500\\
J. Pety, F. Gueth, S. Guillateau, A. Dutrey, 2006, A\&A, \textbf{458}, 841\\
A. F. Pugach, 2004, Astron. Rep \textbf{48}, 470\\
R E. Pudritz, R. Ouyed, C. Fendt, and A. Brandenburg, 2007,
\textit{Protostars and Planets V} (Eds. B. Reipurth,
D. Jewitt, K. Keil, Univ. of Arizona Press, Tucson, 951) p. 277\\
P. N. Safier, 1993a, Astrophys. J. \textbf{408}, 115\\
P. N. Safier, 1993b, Astrophys. J. \textbf{408}, 148\\
V.S. Safronov, 1972, \emph{Evolution of the protoplanetary cloud
and formation of the Earth and planets}, Moscow, Nauka\\
J. Solf and K.H. B\"{o}hm, 1993, Astrophys. J., \textbf{410}, L31\\
H. Shang, A. E. Glassgold, F. H. Shu, and S. Lizano, 2002, Astrophys. J.
\textbf{564}, 853\\
V.S. Shevchenko, K.N. Grankin, M.A. Ibragimov, et al. 1993,
Astrophys. Sp. Sci. \textbf{202}, 121\\
F. Shu, J. Najita, E. Ostriker, et al., 1994, Astrophys. J. \textbf{429}, 781\\
K.R. Stapelfeldt, A.M.Watson, J.E.Krist et al., 1999, ApJ, 516, L95\\
L. V. Tambovtseva, V. P. Grinin, G. Weigelt, 2006, Astron.
Astrophys., \textbf{448}, 633\\
M. Wardle and A. K\"{o}nigl, 1993, Astrophys. J. \textbf{410}, 218\\
A.M. Watson and K. R. Stapelfeldt, 2004, Astrophys. J. \textbf{602}, 860\\
A.M. Watson and K.R. Stapelfeldt, 2007, Astron. J., \textbf{133}, 845\\
S.J. Weidenschilling, 2000, Space Sci. Rev. \textbf{92}, 281\\
J. Woitas, F. Bacciotti, T. P. Ray et al., 2005, Astron. Astrophys. \textbf{432}, 149\\
K. Wood, S.J. Wolk, K.Z.Stanek et al., 2001, Astrophys. J., \textbf{542}, L21\\
K. Wood, and B. Whitney, 1998, Astrophys. J., \textbf{506}, L43\\
\end{document}